\documentclass[journal]{IEEEtran}

\usepackage{cite}

\ifCLASSINFOpdf

\else

\fi

\usepackage{amsmath}

\usepackage{array,multirow}
\usepackage[makeroom]{cancel}
\usepackage{graphicx}
\usepackage{makecell}
\usepackage{array}

\begin{document}

\raggedbottom

\title{Performance Analysis and Enhancements for In-Band Full-Duplex Wireless Local Area Networks}

\author{Murad~Murad,~\IEEEmembership{Student Member,~IEEE,} and~Ahmed~M.~Eltawil,~\IEEEmembership{Senior Member,~IEEE}
\thanks{The authors are with the Department of Electrical Engineering and Computer
Science at the University of California, Irvine, CA 92697 USA (e-mail:
mmurad@uci.edu; aeltawil@uci.edu).}
\thanks{Manuscript submitted January 30 2019.}
\thanks{This work has been submitted to the IEEE for possible publication.
Copyright may be transferred without notice, after which this version may
no longer be accessible.}}

\markboth{}%
{Murad and Eltawil: Performance Analysis and Enhancements for In-Band Full-Duplex Wireless Local Area Networks}

\maketitle

\begin{abstract}
In-Band Full-Duplex (IBFD) is a technique that enables a wireless node to simultaneously transmit a signal and receive another on the same assigned frequency. Thus, IBFD wireless systems can provide up to twice the channel capacity compared to conventional Half-Duplex (HD) systems. In order to study the feasibility of IBFD networks, reliable models are needed to capture anticipated benefits of IBFD above the physical layer (PHY). In this paper, an accurate analytical model based on Discrete-Time Markov Chain (DTMC) analysis for IEEE 802.11 Distributed Coordination Function (DCF) with IBFD capabilities is proposed. The model captures all parameters necessary to calculate important performance metrics which quantify enhancements introduced as a result of IBFD solutions. Additionally, two frame aggregation schemes for Wireless Local Area Networks (WLANs) with IBFD features are proposed to increase the efficiency of data transmission. Matching analytical and simulation results with less than 1\% average errors confirm that the proposed frame aggregation schemes further improve the overall throughput by up to 24\% and reduce latency by up to 47\% in practical IBFD-WLANs. More importantly, the results assert that IBFD transmission can only reduce latency to a suboptimal point in WLANs, but frame aggregation is necessary to minimize it. 
\end{abstract}


\IEEEpeerreviewmaketitle

\section{Introduction}
\IEEEPARstart{T}{he} growth of video traffic led to larger data loads in the downlink (DL) direction to users as compared to uplink (UL) data from users. Even with the prevalence of social networks, where users frequently upload content, the degree of viewership of video has continued to outpace upload leading to a pattern of asymmetric data traffic that is expected to continue for the upcoming years \cite{cisco17a}. Additionally, while Ethernet traffic is declining, WiFi traffic is growing \cite{cisco17b}. Therefore, data traffic in Wireless  Local  Area  Networks (WLANs) is becoming more asymmetric, and this pattern of asymmetry is expected to continue to be the norm. IEEE 802.11 standard defined in \cite{80211} enables client Stations (STAs) to communicate with an Access Point (AP). IEEE 802.11 Distributed Coordination Function (DCF) constitutes the foundation of the Medium Access Control (MAC) protocol for WLANs. By design, IEEE 802.11 DCF does not consider the amount of traffic a node has when facilitating access to the wireless channel. Thus, all WLAN nodes (i.e. the AP and client STAs) have an equal opportunity to access the channel despite the asymmetry between traffic loads in the UL and DL directions. Consequently, traffic asymmetry coupled with equal access to the channel leads to serving data traffic inefficiently in contemporary WLANs. As a result, there is increasing pressure to design future wireless networks that can cope with demands for higher data rates, lower latency, and efficient utilization of resources. Contemporary wireless communications systems are approaching performance limits set by classical analyses. Therefore, there is a need to innovatively design wireless systems that revolutionize the current perception of theoretical limits.

Current WLANs are Half-Duplex (HD), in that they allow either UL or DL transmission over a channel at any given time. A powerful technique is to use In-Band Full-Duplex (IBFD) in order to make an efficient use of the wireless channel. IBFD communications, enabled by Self-Interference Cancellation (SIC) solutions, can theoretically double channel capacity by allowing each wireless node to transmit and receive at the same time and over the same frequency band (see \cite{Sabharwal14}, \cite{Kim15}, or \cite{Zhang16} for a comprehensive coverage of IBFD communications).

Prior advancements in SIC affirm that IBFD is possible, and the legacy assumption of a single transmission over a frequency is no longer a necessity. SIC can be implemented at different levels and in numerous ways. A possible SIC solution can purely take place in the analog domain at both the transmitter and receiver sides \cite{Bharadia13}. On the other hand, SIC can be treated in the digital domain at the transceiver like in \cite{Ahmed13}. An innovative method can extract the Self-Interference (SI) signal from the analog domain and cancel it at the digital domain like in \cite{Ahmed15}. Alternatively, the SI signal can be extracted from the digital domain and cancelled at the analog domain to enable proper reception for the Signal of Interest (SOI)\cite{Duarte12}. A combination of SIC techniques at various levels is often necessary to reduce SI below the noise floor in order to make the received SOI decodable. As a result, IBFD platforms utilizing IEEE 802.11 standard are used to establish operational WLANs (see \cite{Ahmed1502}, \cite{Duarte14}, and \cite{Jain11}).

There were several previously published research attempts to provide analytical models for IBFD MAC protocols. In \cite{Luvisotto18}, a MAC protocol for wireless ad hoc networks is studied based on a three-dimensional Discrete-Time Markov Chain (DTMC) model, but the proposed protocol deviates from IEEE 802.11 DCF mechanism and neglects to derive IBFD-compatible expressions for the probability of transmission. The two-dimensional DTMC model outlined in \cite{Zuo17} does not account for starting a new contention cycle after a node successfully gets an IBFD transmission opportunity, and the model does not follow IEEE 802.11 when it comes to an unsuccessful transmission at the maximum backoff stage. The IBFD MAC protocol in \cite{Liao1502} focuses on simultaneous transmitting and sensing, but the analysis does not fully exploit IBFD benefits for increasing throughput and reducing latency. The authors of \cite{Marlali17} model a new MAC protocol as a three-dimensional DTMC to use IBFD-synchronized transmission only after a successful HD transmission, but the model does not treat collisions accurately. While \cite{Askari14} addresses both throughput and delay in the three-dimensional DTMC analysis for a proposed distributed MAC protocol, the work primarily focuses on multi-hop networks. The IBFD MAC protocol proposed in \cite{Tang15} limits IBFD capabilities to the AP only and substantially neglects to show the details of the theoretical work leading to a basic expression for the  probability of transmission. An Embedded Markov Chain model is used in \cite{Doost16} to study a Carrier-Sense Multiple Access with Collision Avoidance (CSMA/CA) MAC protocol, but the proposed protocol uses a fixed contention window and does not follow IEEE 802.11 Binary Exponential Backoff in case of a collision. The reported results for throughput and delay in \cite{Lee17} show major mismatch between theoretical and simulation results, and the authors state that there was not enough space to include derived analytical expressions for delay calculations. Both inaccuracy of results and lack of a full model for IBFD-WLANs are resolved in this paper. While references \cite{Hu18,Zuo16,Yang15,Liao15,RLiao15} all address IBFD MAC solutions, none of them look into producing an analytical expression for the probability of transmission, which is a substantial part of this paper's contributions.

\begin{figure}[!t]
\centering
\includegraphics[width=\columnwidth,trim=3.0in 0.5in 1.7in 0.9in,clip=true]
{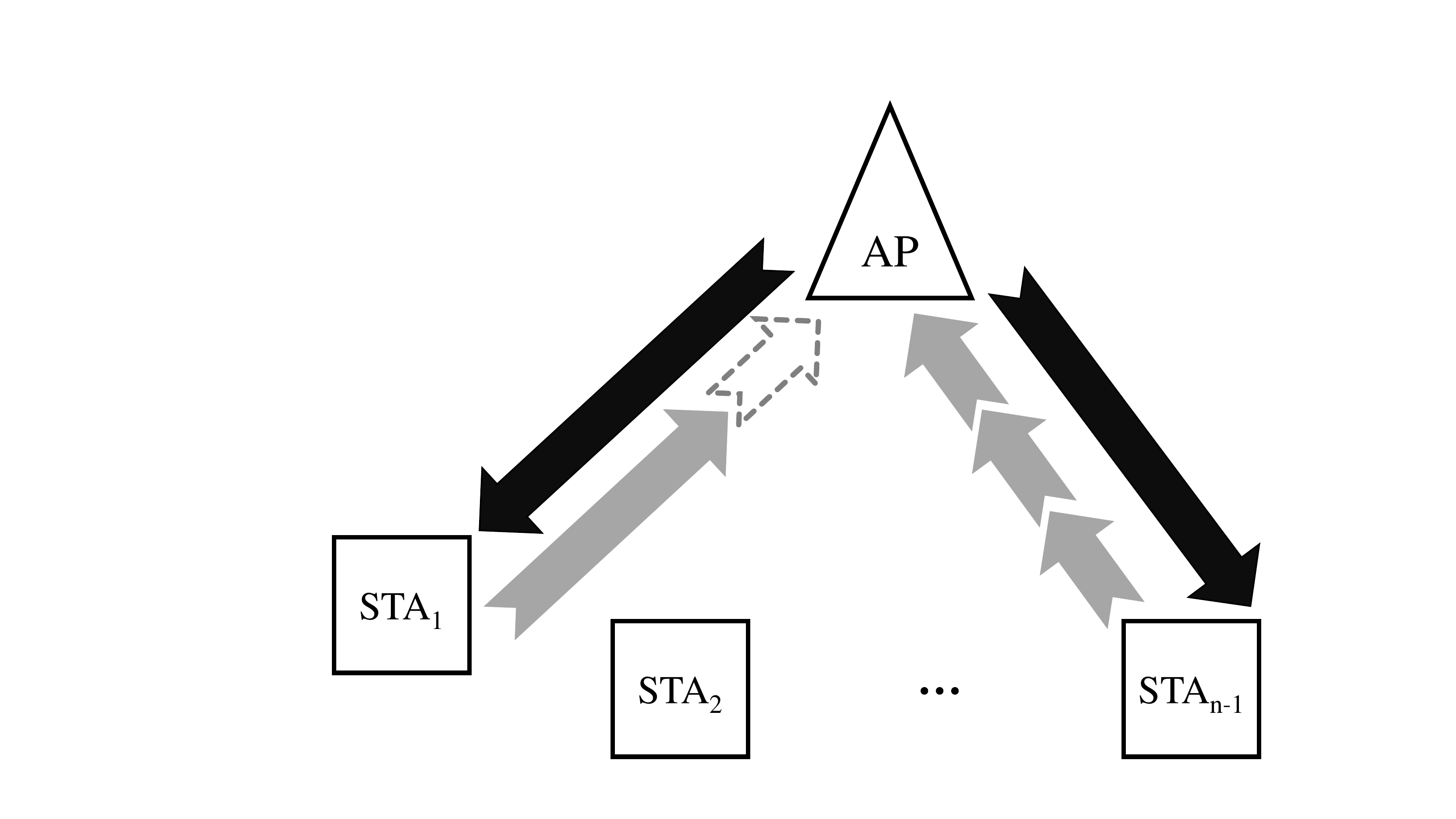}
\caption{A typical IBFD-WLAN with asymmetric traffic loads.}
\label{murad1}
\end{figure}

In this paper, two-dimensional DTMC analysis is used to produce an accurate analytical model for IBFD-WLANs based on IEEE 802.11 DCF. In addition, to serve data traffic even more efficiently, two distributed aggregation solutions for IBFD-WLANs are proposed. Each STA can make an independent decision about the possibility and amount of aggregation based on knowing the size of the traffic it receives. Simulation results indicate that IBFD aggregation leads to both maximizing the throughput of the system and minimizing the average latency of frame delivery. The final contribution of the paper is to formalize metrics in order to study the increase in efficiency that IBFD provides for WLANs. Fig. \ref{murad1} illustrates a typical IBFD-WLAN network with asymmetric traffic and the possibility of frame aggregation.

The rest of this paper is organized as follows. Section II provides an overview of a technical background necessary to establish the HD IEEE 802.11 reference model. Section III outlines the system model and assumptions of the work. Section IV details the proposed analytical model for IBFD-WLAN. Section V establishes performance metrics and explains the proposed IBFD aggregation schemes. Section VI illustrates the generated results. The paper is then concluded in Section VII. 

\section{Technical Background on HD IEEE 802.11 DCF}
A well-celebrated analytical model for IEEE 802.11 DCF was presented in \cite{Bianchi00}. This model was subsequently revised a number of times, especially when it comes to the the \textit{probability of transmitting} $(\tau)$. To generate highly accurate results for HD IEEE 802.11 DCF, $\tau$ is adopted from the refined model published in \cite{Tinnirello10} as
\begin{equation}
\tau =  \frac{1}{1+\frac{1-p}{1-p^{R+1}} \sum_{i=0}^{R} p^i (2^iW-1)/2-\frac{1-p}{2}}
\label{tau_HD}
\end{equation}
where \textit{W} is the initial Contention Window (CW$_\textnormal{min}$), $R$ is the maximum number of re-transmission attempts, and $p$ is the \textit{conditional collision probability}. An STA experiences a collision when at least one other wireless node concurrently transmits. Therefore,
\begin{equation}
p = 1-(1-\tau)^{n-1}
\label{p_HD}
\end{equation}
where \textit{n} is the total number of nodes. Equations (\ref{tau_HD}) and (\ref{p_HD}) can simply be solved numerically to calculate the values of $\tau$ and $p$ for each node. 

The \textit{probability of a successful transmission} $(P_s)$ is the conditional probability there is exactly one transmission given there is at least one transmission, which is equal to
\begin{equation}
P_s =\frac{n\tau(1-\tau)^{n-1}}{1-(1-\tau)^n}.
\end{equation}

The \textit{throughput} $(S)$ in Mega bits per second (Mbps) is calculated as
\begin{equation}
S = \frac{P_{s} P_{tr} \overline{E[P]}}{(1-P_{tr})\sigma+P_{tr} P_{s} \overline{T_{s}} + P_{tr}(1-P_{s})\overline{T_{c}}}
\end{equation}
where $P_{tr} $ is the \textit{probability that there is at least a transmission}, and it is given by
\begin{equation}
P_{tr} = 1-(1-\tau)^n.
\end{equation}
According to \cite{Tinnirello10}, an accurate characterization for the throughput is achieved if the \textit{expected payload size} $\overline{E[P]}$, the \textit{expected time needed for a successful transmission} $(\overline{T_s})$, and the \textit{expected time spent during a collision} $(\overline{T_c})$ are respectively expressed as

\begin{equation}
   \overline{E[P]} = E[P] \frac{W}{W-1}
\end{equation}

\begin{equation}
\overline{T_s} = T_s \frac{W}{W-1} + \sigma,
\end{equation}
and
\begin{equation}
\overline{T_c} = T_c + \sigma,
\end{equation}
where
\begin{equation}
T_s = \textnormal{H} + \textnormal{payload time} + \textnormal{SIFS} + \textnormal{ACK} +\textnormal{DIFS}
\label{Ts}
\end{equation}
and
\begin{equation}
T_c = \textnormal{H} + \textnormal{collision time} +\textnormal{SIFS}+\textnormal{ACK}+\textnormal{DIFS}.
\label{Tc}
\end{equation}
H is the total time for both PHY and MAC headers. Values for headers, SIFS, ACK, DIFS, and time slot duration $(\sigma)$ are set by IEEE 802.11 standard. Table \ref{murad.t1} shows the values of PHY and MAC parameters based on the IEEE 802.11ac release \cite{80211ac}.
\begin{table} [!b]
\caption{IEEE 802.11ac PHY and MAC Parameters.}
\label{murad.t1}
\centering
\begin{tabular}{l|r}
Parameter & Value \\\hline
Frequency & 5 GHz band\\
Channel bandwidth & 80 MHz \\
Modulation scheme & 16-QAM \\
Code rate & 1/2 \\
Spatial streams & 2$\times$2 MIMO\\
PHY header duration & 44 $\mu$s \\
Guard Interval (GI) & 800 $n$s \\
Transmission rate & 234 Mbps\\
Basic rate & 24 Mbps\\
MAC header size & 36 bytes \\
FCS size & 4 bytes \\
ACK size & 14 bytes \\
MPDU$_{\text{max}}$  size & 7,991 bytes\\
Slot duration ($\sigma$) & 9 $\mu$s \\
SIFS duration & 16 $\mu$s \\
DIFS duration & 34 $\mu$s \\
CW$_{\text{min}}$ & 16 \\
CW$_{\text{max}}$ & 1024
\end{tabular}
\end{table}

Considering the system model detailed in the next section, analytical expressions for the expected size of successfully transmitted MAC Protocol Data Unit (MPDU) and the expected size of a collision are thoroughly derived in \cite{Murad18} and can respectively be simplified as
\begin{equation}
(E[P])^\textit{\tiny HD} = \frac{n+1}{2n}\cdot{\text{MPDU}_{\text{max}}}
\end{equation}
and
\begin{equation}
\begin{split}
&(E[P^*])^\textit{\tiny HD} \approx\\
&\Bigg[\frac{0.3519\times\tau(1-(1-\tau)^{n-1})}{1-(1-\tau)^n-n \tau(1-\tau)^{n-1}} + 0.6481 \Bigg]{\text{MPDU}_{\text{max}}}.
\end{split}
\end{equation}

Latency is calculated as the average time from the instant a frame becomes Head-of-Line (HOL) until the frame is successfully delivered. The analytical expression for latency in HD IEEE 802.11 is derived in \cite{Bianchi08} directly from the well-known Little's Theorem (see \cite{Gallager92} for further explanation) as
\begin{equation}
    D = \frac{n}{S/E[P]}.
\label{D_HD}
\end{equation}

\section{System Model}
This paper assumes a WLAN with an AP and $n-1$ client STAs using IEEE 802.11ac standard to communicate over a single channel. The basic mode of DCF without Request to Send/Clear to Send (RTS/CTS) handshake is assumed. In case of a collision, total frame loss occurs (no capture effect). Error-free PHY transmission is assumed, and all nodes can detect one another (no hidden terminals). The AP always has a load of MPDU$_{\text{max}}$. A saturated buffer at each node is assumed (i.e. there is always traffic to transmit), and frame aggregation is possible by combining more than one MPDU to make a larger aggregated MPDU. Client STAs have Symmetry Ratio (SR) values
where the value of SR at STA$_{i}$, indicated here as $\rho_i$, is defined in \cite{Murad17} as the ratio of the UL load over the DL load. If the traffic load is designated as ($L$) and transmission time as ($T$), then $\rho$ is
\begin{equation}
\rho \overset{\Delta}{=} \frac{L_{UL}}{L_{DL}} = \frac{T_{UL}}{T_{DL}}.
\end{equation}
Each STA$_i$ has 0.1 $\leq \rho_i$ $\leq$ 0.9. STAs keep their original $\rho$ values constant throughout each simulation run.

In IBFD-WLAN scenarios, all transmissions occur as IBFD between the AP and an STA. The Full-Duplex Factor (FDF) defined in \cite{Murad18} as the average of all $\rho$ values of the client STAs in the network can be calculated as
\begin{equation}
\Phi \overset{\Delta}{=} \frac{\sum_{i=1}^{n-1} \rho_i}{n-1}.
\end{equation}
For an HD system, $\Phi = 0$.

\section{Analytical Model for IBFD-WLAN}
Prior work to model contemporary IEEE 802.11 DCF was based on the assumption that all transmission takes place in a Time-Division Duplexing (TDD) fashion, which is an HD scheme. Therefore, the prominent model originally presented in \cite{Bianchi00} is no longer valid when the transmission is IBFD. In HD systems, two key parameters, namely the probability of transmitting ($\tau$) and the conditional collision probability ($p$), determine the performance of a WLAN at the MAC sub-layer. However, when the system is IBFD, both $\tau$ and $p$ must be revised. First, unlike the HD case where AP and STAs share equal $\tau$ and $p$ values, there are $\tau_{_{AP}}$, $\tau_{_{STA}}$, $p_{_{AP}}$, and $p_{_{STA}}$ values in an IBFD network. Second, in addition to the \textit{direct} transmission probability, $\tau$, which happens when a node wins the contention for the channel, there is potential for IBFD \textit{reply-back} transmission when the node is not in direct transmission ($\overline{\tau}$). The probability of reply-back transmission $(\beta)$ for a tagged node happens when another node is in direct transmission with the tagged node. Third, collisions are treated differently for IBFD systems compared to a contemporary HD WLAN, which is thoroughly explained in \cite{Murad18}.

In this section, analytical work is carried out to construct a model for IBFD-WLAN based on IEEE 802.11 DCF protocol. Key parameters needed to calculate important performance metrics are defined. All parameters take into account IBFD and its effects on the behavior of wireless nodes at MAC-level operations.

\subsection{Revised Probability of Transmission $(\tau)$}
\begin{figure*}[ht]
  \includegraphics[width=\textwidth,height=10cm]{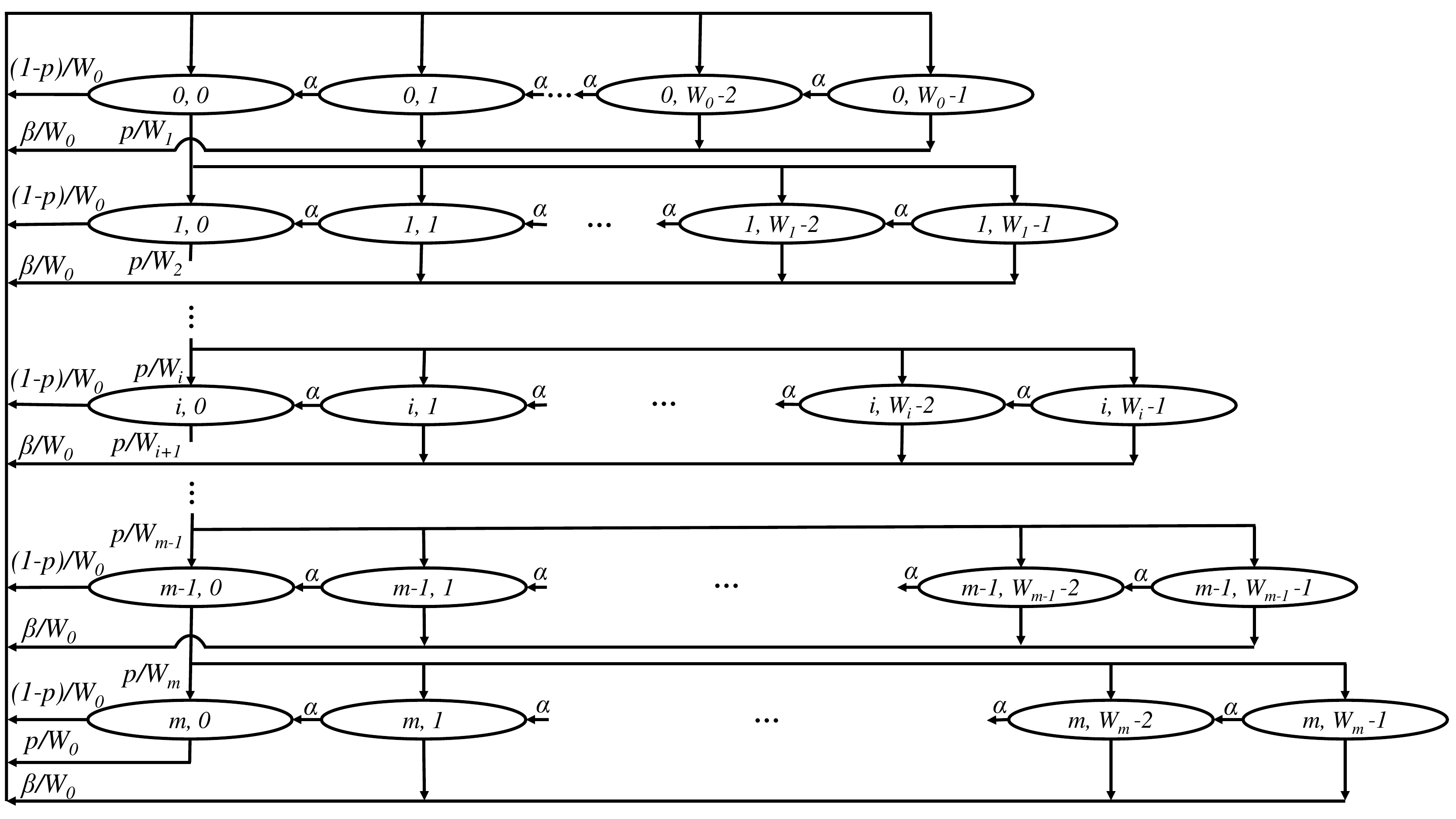}
  \caption{Two-dimensional DTMC representing backoff stage and backoff counter for each wireless node.}
  \label{murad2}
\end{figure*}

Fig. \ref{murad2} shows the model adopted for IBFD-WLAN. The two-dimensional DTMC model represents each state in terms of backoff stage, $i$, and backoff counter, $k$. Unlike \cite{Lee17}, the model in Fig. \ref{murad2} resets its backoff stage to zero if a frame experiences a collision while the transmitting node is at the maximum backoff stage $m$. 

The transition probabilities for the DTMC model are as follows
\begin{align}
&P\{0,k_0|i,0\} = \frac{1-p}{W_0} &&i\in[0,m-1],\\
& &&k_0\in[0,W_0-1] \nonumber\\
&P\{i,k|i-1,0\} = \frac{p}{W_i} && i\in[1,m],\\
& &&k\in[0,W_i-1] \nonumber\\
&P\{0,k_0|m,0\} = \frac{1}{W_0} && k_0\in[0,W_0-1]\\
&P\{i,k-1|i,k\} = \alpha && i\in[0,m],\\
& &&k\in[1,W_i-1] \nonumber\\
&P\{0,k_0|i,k\} = \frac{\beta}{W_0} && i\in[0,m],\\
& &&k\in[1,W_i-1],\nonumber\\
& &&k_0\in[0,W_0-1].\nonumber
\end{align}

The stationary distribution of the chain is represented as
\begin{align}
&b_{i,k} \overset{\Delta}{=}\lim_{t\to\infty} P\{s(t)=i, b(t)=k\} &&i\in[0,m],\\
& &&k\in[0,W_i-1]\nonumber
\end{align}
where $s(t)$ and $b(t)$ are respectively the stochastic processes for the backoff stage and backoff counter as in \cite{Bianchi00}.

Direct transmission happens when a node is at any of the possible $b_{i,0}$ states. Therefore, the probability of direct transmission is
\begin{equation}
\tau \overset{\Delta}{=} \sum_{i=0}^{m} b_{i,0}.
\end{equation}

By applying the normalization condition, the following result can be directly obtained
\begin{equation}
\begin{split}
1 &= \sum_{i=0}^{m} \sum_{k=0}^{W_i - 1} b_{i,k} = \sum_{i=0}^{m} b_{i,0} + \sum_{i=0}^{m} \sum_{k=1}^{W_i - 1} b_{i,k}\\ &= \tau + \sum_{i=0}^{m} \sum_{k=1}^{W_i - 1} b_{i,k}\\ &\Rightarrow \overline{\tau} \overset{\Delta}{=} 1 - \tau = \sum_{i=0}^{m} \sum_{k=1}^{W_i - 1} b_{i,k}
\end{split}
\end{equation}
where $\overline{\tau}$ is the probability that there is potential for IBFD reply-back transmission when a node is not in direct transmission.

The expressions for both $b_{i,0}$ and $b_{0,0}$ are respectively as follows (see Appendix A for complete derivations)
\begin{equation}
b_{i,0} = b_{0,0} (\frac{p}{1-\alpha})^i \prod_{j=1}^{i} \frac{1-\alpha^{W_j}}{W_j}, 1 \leq i \leq m
\end{equation}
and
\begin{equation}
b_{0,0} = \frac{\frac{1-\alpha^{W_0}}{W_0}\Bigg[\frac{\alpha - p}{1- \alpha} \cdot \tau + 1\Bigg]}{1-(\frac{p}{1-\alpha})^{m+1} \displaystyle \prod_{j=0}^{m} \frac{1-\alpha^{W_j}}{W_j}}.
\label{b0_0}
\end{equation}
Thus, $\tau$ is readily calculated as
\begin{equation}
\begin{split}
\tau &= \sum_{i=0}^{m} b_{i,0} = b_{0,0} + \sum_{i=1}^{m} b_{i,0}\\
&= b_{0,0}\Bigg[1+\sum_{i=1}^{m} (\frac{p}{1-\alpha})^i \prod_{j=1}^{i} \frac{1-\alpha^{W_j}}{W_j}\Bigg].\\
\end{split}
\label{tau}
\end{equation}

Given that calculating $\alpha$ and $p$ is treated in the next two sub-sections, the value of $\tau$ can be numerically calculated using (\ref{b0_0}) and (\ref{tau}). While $\tau$ is the same for the AP and STAs in contemporary HD IEEE 802.11 DCF, its value in an IBFD-WLAN is different depending on if the transmitting node is the AP or an STA. $\tau_{_{AP}}$ and $\tau_{_{STA}}$ are calculated based on the corresponding $\alpha_{_{AP}}$, $p_{_{AP}}$, $\alpha_{_{STA}}$, and $p_{_{STA}}$ values. Finally, the average probability of transmission in the network is calculated as
\begin{equation}
    \tau^\textit{\tiny IBFD} = \frac{1}{n} \tau_{_{AP}} + \frac{n-1}{n} \tau_{_{STA}}.
\end{equation}

\subsection{Probability of reply-back IBFD transmission $(\beta)$}
In IBFD-WLAN, each node has an opportunity for indirect transmission. Whenever a node is not in any of the states represented by $\tau$, the node can have the opportunity to transmit if another node is transmitting to it. There are two cases for this to happen as follows
\begin{enumerate}
\item When the AP is silent, it still has an opportunity to transmit whenever an STA is transmitting. This probability can be represented as
\begin{equation}
\begin{split}
\beta_{_{AP}} &= (n-1) \tau_{_{STA}} (1-\tau_{_{STA}})^{n-2}\\\\
&= (n-1) \tau_{_{STA}} (\overline{\tau}_{_{STA}}
)^{n-2}.
\end{split}
\label{beta_AP}
\end{equation}
\item When an STA is silent, it still has an opportunity to transmit whenever the AP is transmitting to that particular STA. This probability is represented as
\begin{equation}
\begin{split}
\beta_{_{STA}} &= \frac{\tau_{_{AP}}(1-\tau_{_{STA}})^{n-2}}{n-1}\\\\
&= \frac{\tau_{_{AP}}(\overline{\tau}_{_{STA}})^{n-2}}{n-1}.
\end{split}
\label{beta_STA}
\end{equation}
\end{enumerate}
For the special case when $n=2$, (\ref{beta_AP}) and (\ref{beta_STA}) respectively become $\beta_{_{AP}} = \tau_{_{STA}}$ and $\beta_{_{STA}} = \tau_{_{AP}}$, which is compatible with the intuition that the AP has a reply-back opportunity whenever the STA is transmitting and vice versa. Also, based on Fig. \ref{murad2}, the following equation can be used to calculate both $\alpha_{_{AP}}$ and $\alpha_{_{STA}}$ according to the corresponding $\beta_{_{AP}}$ and $\beta_{_{STA}}$ values 
\begin{equation}
\alpha = 1 - \beta.
\end{equation}

\subsection{Revised Conditional Collision Probability $(p)$}
In HD IEEE 802.11 networks, collisions are treated in the same way for the AP and all STAs. As a result, conditional collision probability, $p$, is defined in \cite{Bianchi00} as previously stated (see (\ref{p_HD}) in section II). However, in IBFD scenarios, the conditional collision probability for the AP $(p_{_{AP}})$ is different from that of an STA $(p_{_{STA}})$. For the AP, collision-free transmission happens in either of the following two cases
\begin{enumerate}
\item The AP is in direct transmission while all STAs are silent.
\item The AP is in direct transmission with a tagged STA, and this tagged STA is directly transmitting back to the AP while all other STAs are silent. 
\end{enumerate}
Therefore, the conditional collision probability for the AP can be expressed as
\begin{equation}
\begin{split}
p_{_{AP}} &= 1 - \Bigg[(1-\tau_{_{STA}})^{n-1} + \tau_{_{STA}} (1-\tau_{_{STA}})^{n-2}\Bigg]\\
 &= 1 - \Bigg[(\overline{\tau}_{_{STA}})^{n-1} + \tau_{_{STA}} (\overline{\tau}_{_{STA}})^{n-2}\Bigg].
\end{split}
\label{p_AP_IBFD}
\end{equation}

For the conditional collision probability of an active STA, transmission without collision takes place when either one of the below scenarios is true
\begin{enumerate}
\item The AP is silent, and so are all other STAs.
\item The AP is directly transmitting back to the active STA while all other STAs are silent. 
\end{enumerate}
Consequently, the conditional collision probability of a tagged STA is
\begin{equation}%
\begin{split}
&p_{_{STA}} \\\\&= 1 - \Bigg[(1-\tau_{_{AP}}) (1-\tau_{_{STA}})^{n-2} + \frac{\tau_{_{AP}} (1-\tau_{_{STA}})^{n-2}}{n-1}\Bigg]\\
&= 1 - \Bigg[\overline{\tau}_{_{AP}} (\overline{\tau}_{_{STA}})^{n-2} + \frac{\tau_{_{AP}} (\overline{\tau}_{_{STA}})^{n-2}}{n-1}\Bigg].
\end{split}
\label{p_STA_IBFD}
\end{equation}

Similar to how it was detailed in \cite{Murad18} with simplified assumptions regarding $\tau$, (\ref{p_AP_IBFD}) and (\ref{p_STA_IBFD}) indicate that a collision-free mode of IBFD transmission is achieved when $n=2$ since both equations respectively evaluate to $p_{_{AP}} = 0$ and $p_{_{STA}} = 0$.

\subsection{Revised Probability of Successful Transmission $(P_s)$}
The probability of successful transmission for an IBFD-WLAN, $P^\textit{\tiny IBFD}_s$, happens during any one of the following four conditional probabilities 
\begin{enumerate}
    \item There is exactly one direct transmission by the AP and all STAs are silent given there is at least a direct transmission.
    \item There is exactly one direct transmission by an STA while the AP and all other STAs are silent given there is at least a direct transmission.
    \item There are exactly two direct transmissions, one is by the AP and the other is by the corresponding STA back to the AP given that there is at least a direct transmission.
    \item There are exactly two direct transmissions, one is by an STA and the other is by the AP back to the STA given that there is at least a direct transmission.
\end{enumerate}
Therefore, $P^\textit{\tiny IBFD}_s$ is expressed as (see Appendix B for complete derivation)
\begin{equation}
\begin{split}
&P^\textit{\tiny IBFD}_s=\\
\\
&\frac{\tau_{_{AP}}(\overline{\tau}_{_{STA}})^{n-1}+(n-1)\tau_{_{STA}}(\overline{\tau}_{_{AP}})(\overline{\tau}_{_{STA}})^{n-2}}{1-[(\overline{\tau}_{_{AP}}) (\overline{\tau}_{_{STA}})^{n-1}]}\\
\\
&+\frac{\tau_{_{AP}}\tau_{_{STA}}(\overline{\tau}_{_{STA}})^{n-2}}{(n-1)\big\{1-[(\overline{\tau}_{_{AP}}) (\overline{\tau}_{_{STA}})^{n-1}]\big\}}.
\end{split}
\label{P_s_IBFD}
\end{equation}
When (\ref{P_s_IBFD}) is evaluated at $n=2$, (\ref{P_s_IBFD}) becomes $P^\textit{\tiny IBFD}_s= 1$ indicating that every transmission is successful, which is consistent with the result assuming $\tau_{_{AP}}=\tau_{_{STA}}=\tau$ reported in \cite{Murad18}. 

\section{IBFD-WLAN System Performance Metrics}
In this section, IBFD-compatible metrics are outlined. The purpose of composing a portfolio of metrics is to measure the enhancements added by IBFD to WLAN performance. The metrics will be used to generate the results in Section VI.

\subsection{Network Throughput}
Throughput gain by IBFD was previously addressed in the published conference paper \cite{Murad18}. However, the focus there was primarily on presenting how collisions are treated and consequently affect the performance of normalized aggregate throughput. Therefore, the value of $\tau$ was directly obtained from \cite{Bianchi00}. The analysis in \cite{Murad18} is revisited here to include a more accurate model that considers both $\tau_{_{AP}}$ and $\tau_{_{STA}}$ derived in this paper. Additionally, the total network throughput is calculated in the absolute sense in terms of Mbps instead of the normalized value. Therefore, the \textit{network throughput} can be expressed as
\begin{equation}
S^\textit{\tiny IBFD} = \frac{P^\textit{\tiny IBFD}_s P^\textit{\tiny IBFD}_{tr} E[P] (1+\Phi)}{\overline{P^\textit{\tiny IBFD}_{tr}} \sigma+P^\textit{\tiny IBFD}_{tr} P^\textit{\tiny IBFD}_s T_{s} + P^\textit{\tiny IBFD}_{tr}\overline{P^\textit{\tiny IBFD}_s} T_{c}}
\end{equation}
where the probability of transmitting is revised here to include both $\tau_{_{AP}}$ and $\tau_{_{STA}}$ as
\begin{equation}
\begin{split}
P^\textit{\tiny IBFD}_{tr} &= 1-[(1-\tau_{_{AP}}) (1-\tau_{_{STA}})^{n-1}]\\
&= 1-[(\overline{\tau}_{_{AP}}) (\overline{\tau}_{_{STA}})^{n-1}].
\end{split}
\label{P_tr_IBFD}
\end{equation}
Note that
\begin{equation}
\overline{P^\textit{\tiny IBFD}_{tr}} = 1 - P^\textit{\tiny IBFD}_{tr}
\end{equation}
and
\begin{equation}
\overline{P^\textit{\tiny IBFD}_s} = 1 - P^\textit{\tiny IBFD}_s.
\end{equation}
Both $T_s$ and $T_c$ are reported respectively as (\ref{Ts}) and (\ref{Tc}) in section II. As explained in \cite{Murad18}, since UL $<$ DL, both the expected  size  of  successfully transmitted MPDU and the expected size of a collision are equal to the load of the AP as follows
\begin{equation}
(E[P])^\textit{\tiny IBFD} = (E[P^*])^\textit{\tiny IBFD} = E[P_{_{AP}}] = {\text{MPDU}_{\text{max}}}.
\end{equation}

\subsection{Frame Aggregation}
Frame aggregation at the MAC sub-layer was introduced in the legacy IEEE 802.11n release \cite{80211n} as a way to increase the efficiency of utilizing the channel by reducing the overhead. Aggregation enables an STA to concatenate several frames into a single transmission. As a result, the overhead is reduced since there is no need to allocate transmission time for more than a single header duration if the frames are combined into one transmission. Frame aggregation can be either Aggregated MAC Service Data Unit (A-MSDU) or Aggregated MPDU (A-MPDU). For the sake of this paper, MPDUs are aggregated. Details about frame aggregation in IEEE 802.11 standard can be found in \cite{Karmakar17} and \cite{Perahia13}.

\begin{figure}[!b]
\centering
\includegraphics[width=\columnwidth,trim=4in 1.4in 3.5in 1.1in,clip=true]{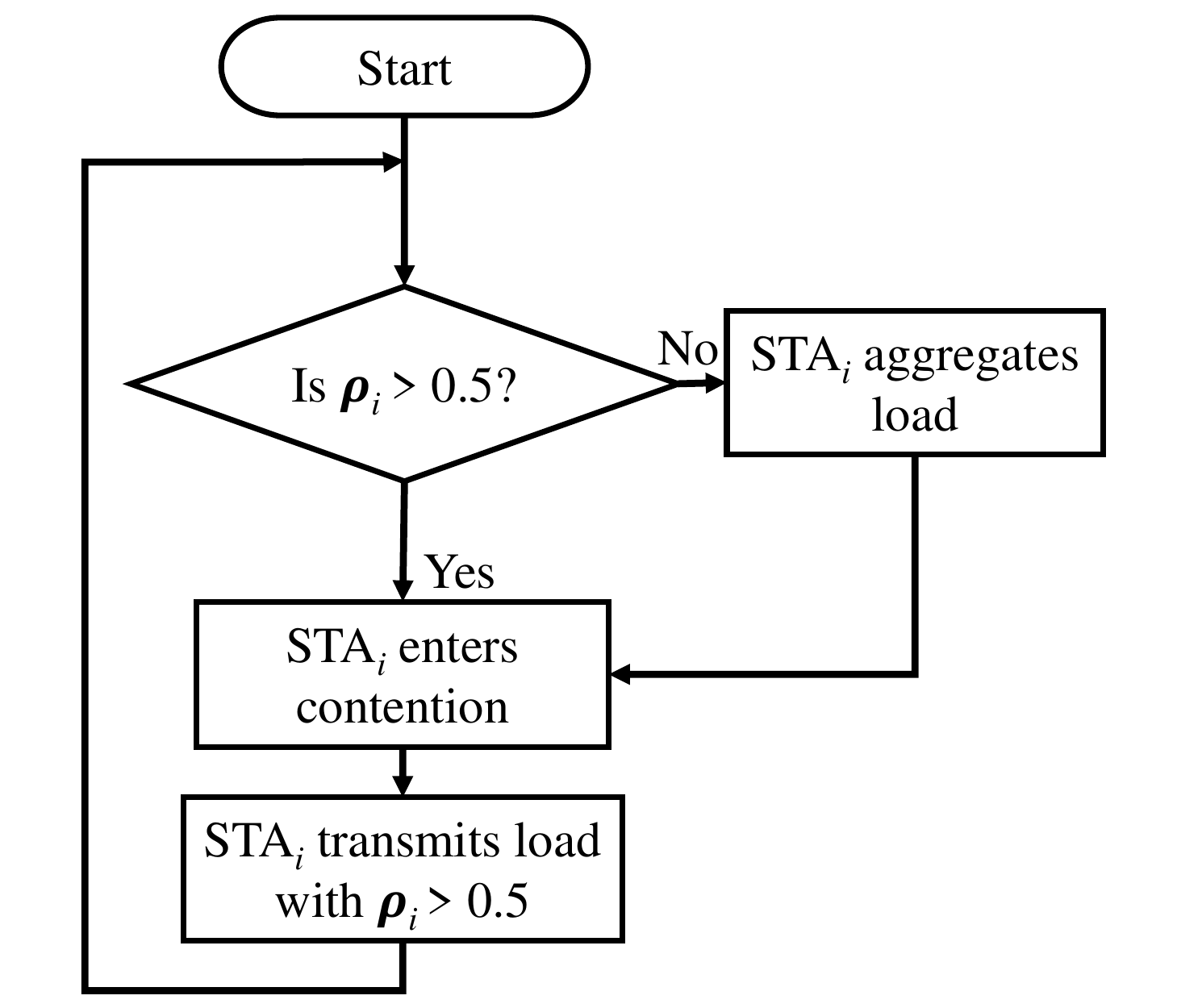}
\caption{Flow chart of the proposed aggregation schemes}
\label{murad3}
\end{figure}

The goal of frame aggregation in this paper is to increase traffic symmetry between UL and DL data loads and consequently serve traffic more efficiently. This paper introduces two aggregation schemes which significantly improve network throughput, average latency, and link utilization. Fig. \ref{murad3} shows a flow chart of how aggregation is performed based on the value of $\rho$ at each STA. The frame aggregation schemes are namely \textit{Dual-Frame Aggregation} and \textit{Multi-Frame Aggregation}.

\subsubsection{Dual-Frame Aggregation}
in this aggregation scheme, any STA with $\rho \leq 0.5$ doubles its transmission load by aggregating two MPDUs. Thus,
\begin{equation}
\rho^{\textit{\tiny dual}}_{\textit{\tiny new}} := 2 \times \rho_{\textit{\tiny current}}.
\end{equation}
The \textit{IBFD aggregation factor} (i.e. number of aggregated frames) for dual-frame aggregation is $\gamma^{\textit{\tiny dual}} = 2$ if aggregation takes place. In this case, it is still guaranteed that each STA can fit its transmission while the IBFD connection is established with the AP since UL $\leq$ DL even after frame aggregation. Dual-frame aggregation increases the utilization of the available UL transmission time that would be otherwise not used.

\subsubsection{Multi-Frame Aggregation}
some STAs with $\rho < 0.5$ can aggregate more than two frames in a transmission. The following is a simple rule to calculate IBFD aggregation factor for multi-frame aggregation in order to determine how many frames each STA can aggregate based on its current $\rho$ value
\begin{equation}
\gamma^{\textit{\tiny multi}} \overset{\Delta}{=} \texttt{floor}\bigg(\frac{1}{\rho}\bigg).
\end{equation}
In this aggregation technique, 
\begin{equation}
\rho^{\textit{\tiny multi}}_{\textit{\tiny new}} := \texttt{floor}\bigg(\frac{1}{\rho_{\textit{\tiny current}}}\bigg) \times \rho_{\textit{\tiny current}}.
\end{equation}
The result of multi-frame aggregation is that STAs with very small $\rho$ values can aggregate several frames, which increases both UL/DL traffic symmetry and utilization of available UL transmission time. For any STA with $\rho>0.5$, no aggregation takes place ($\gamma^{\textit{\tiny dual}} = \gamma^{\textit{\tiny multi}} = 1$).

\subsection{Average Latency}
A similar analysis to the work in \cite{Bianchi08} is used to derive an analytical expression for average latency in IBFD-WLAN. According to \cite{Gallager92}, Little's Theorem classically states that the average number of customers in a system $(N)$ is equal to the average arrival rate of the customers $(\lambda)$ multiplied by the average delay per customer in the system $(T)$. Thus,
\begin{equation}
    N = \lambda T \Rightarrow T = \frac{N}{\lambda}.
\end{equation}

Since saturated traffic is assumed in the network, the number of customers in the system is always equal to the number of nodes $n$. Unlike the case of HD latency considered in section II as (\ref{D_HD}), the case of IBFD transmission is more involved. The expected number of aggregated MPDUs, $E[\gamma]$, must be taken into consideration when calculating the frame arrival rate, which is equal to the frame departure rate since in a saturated buffer, a new (possibly aggregated) frame promptly arrives once the HOL frame is transmitted. Since one frame is transmitted in the DL direction and $E[\gamma]$ frames are transmitted in the UL direction, Little's Theorem can be applied to calculate the average latency per frame in an IBFD-WLAN as follows
\begin{equation}
    \begin{split}
    D^\textit{\tiny IBFD}&=\frac{n}{(1+E[\gamma])\cdot S^\textit{\tiny IBFD}\bigg/\Big[E[P_{_{AP}}]\cdot(1+\Phi)\Big]}\\ \\
    &=\frac{n \cdot E[P_{_{AP}}]\cdot(1+\Phi)}{(1+E[\gamma])\cdot S^\textit{\tiny IBFD}}.
    \end{split}
\end{equation}

\subsection{IBFD Link Utilization}
Throughput quantifies successful transmission of data over the total time including successful, collision, and sensing durations while considering added overhead. A metric that is worth introducing is IBFD link utilization, $\eta$, in order to quantify the efficiency of using the link (in both UL and DL directions) for transmission of useful data loads without the overhead. Since UL transmission is less than or equal to DL transmission, IBFD link utilization can be defined as
\begin{equation}
\eta \overset{\Delta}{=} \frac{1+\Phi}{2} \times 100 \%.
\end{equation}
Ideally, if the channel is fully utilized in both UL and DL directions (i.e. $\rho$ = 1 at each STA $\Rightarrow \Phi=1$), then $\eta = 100\%$ indicating a fully utilized and symmetrical link. IBFD link utilization is particularly crucial when assessing the benefits of frame aggregation, and this becomes clear by the numerical results reported in the next section. 

\section{Results and Evaluation}
In order to confirm the validity of the analytical model detailed in this paper for IBFD-WLAN, results based on simulated IEEE 802.11ac standard are used as a baseline. Analytical and simulation results for both network throughput and average latency in standard HD IEEE 802.11, IBFD-WLAN, IBFD-WLAN with dual-frame aggregation, and IBFD-WLAN with multi-frame aggregation are presented. In all generated results, the IBFD-WLAN analytical model provides values that closely match the simulated results within 1\% error or less. Throughput quantifies successfully transmitted data over the total time. Latency quantifies the average time needed to successfully deliver an MPDU frame from the time the frame becomes HOL until an ACK frame is received. IBFD link utilization is used as a new metric to measure the enhancements added by IBFD aggregation techniques. 

Three sets of results are provided. First, both the IBFD-WLAN model proposed in this paper and the analytical model published in \cite{Lee17} are compared to simulated results. Then, the performance of the network in terms of throughput, latency, and utilization when $\rho$ values are deterministic is evaluated in order to illustrate the aggregation schemes and their benefits. Finally, the performance results are repeated when $\rho$ values are random to show a more practical scenario for a typical network.

\subsection{Accuracy of the proposed IBFD-WLAN model}
\begin{figure}[!t]
\centering
\includegraphics[width=\columnwidth,trim=1.4in 3.3in 1.7in 3.5in,clip=true]
{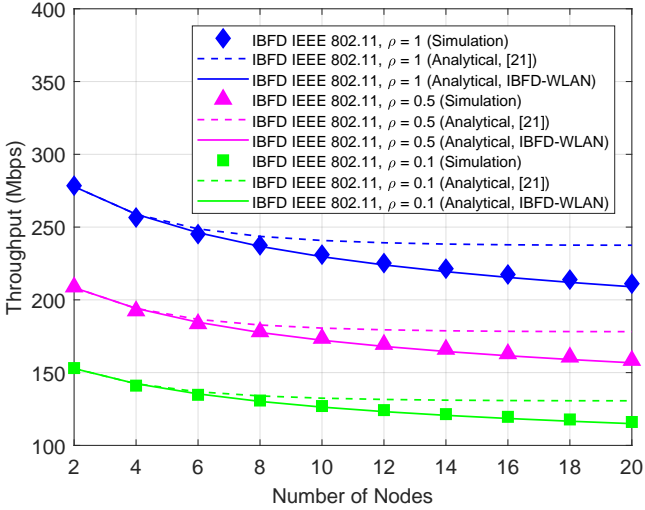}
\caption{Comparison between throughput results from the proposed IBFD-WLAN analytical model and the analytical model published in \cite{Lee17}.}
\label{murad4}
\end{figure}

The reported graphs in \cite{Lee17} show noticeable discrepancies between simulation and analytical results for the performance of the network. Therefore, there is a need for an accurate analytical model that realizes the impact of IBFD on WLANs. Fig. \ref{murad4} shows simulation results for network throughput versus number of nodes $(n)$ in a WLAN based on IEEE 802.11 standard for three deterministic $\rho$ values. The corresponding analytical results based on the IBFD-WLAN framework proposed in this paper are plotted. Additionally, the corresponding cases based on the analytical work reported in \cite{Lee17} are plotted for comparison. To make the comparison fair, the testing of the two analytical models was made closely similar by primarily using different formulas from the corresponding models for $\tau$ calculations while keeping all other parameters identical using the latest IEEE 802.11ac release (which \cite{Lee17} does not originally use). It is clear that at low $n$ values, both analytical models match the simulated results. However, at higher $n$ values, the analytical model proposed in this paper continues to match the simulation results while the model from \cite{Lee17} provides overly optimistic results. For each curve, the average error between the simulated scenario and analytical results based on IBFD-WLAN model is always less than 1\% (matching the accuracy of the well-studied HD IEEE 802.11 model). On the other hand, the mismatch introduced by \cite{Lee17} consistently increases as the number of nodes increases until it reaches about 13\% in all three cases when $n=20$. The high error at high $n$ values cannot be justified by the fact that the model in \cite{Lee17} assumes, unlike IEEE 802.11 standard, infinite re-transmission attempts at the maximum backoff stage until the frame is successfully delivered. This difference alone can only provide much smaller deviation between simulation and analytical cases. The inaccurate results at high $n$ values are also apparent in the reported plots in \cite{Lee17}. Latency comparison is not performed here since no complete analytical model for latency was reported in \cite{Lee17}.

\subsection{Deterministic $\rho$ Values}
\begin{figure}[!b]
\centering
\includegraphics[width=\columnwidth,trim=1.4in 3.3in 1.7in 3.5in,clip=true]
{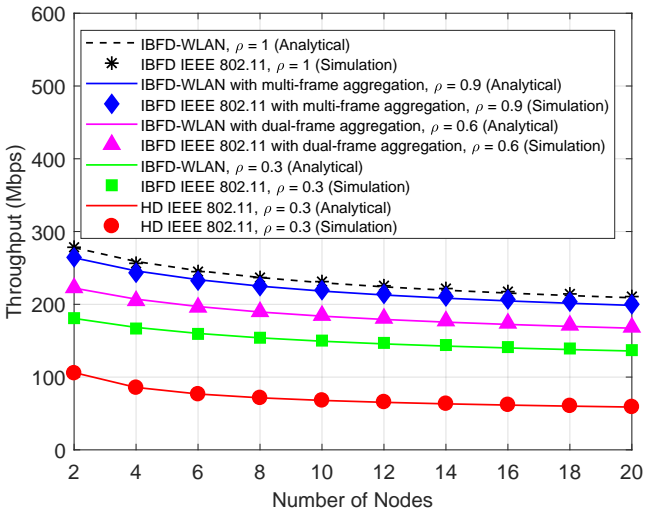}
\caption{Throughput versus number of nodes $(\rho=0.3)$}
\label{murad5}
\end{figure}
Fig. \ref{murad5} shows both analytical and simulation results for network throughput versus the number of nodes when each client STA always has $\rho = 0.3$ originally. The AP always transmits a load of $\text{MPDU}_{\text{max}}$. The case for a standard HD IEEE 802.11 network is shown as a baseline case. When IBFD mode is enabled, the improvement in throughput depends on the number of nodes. For 2 nodes, the throughput increases by 72\% compared to the case of HD IEEE 802.11 protocol. For 20 nodes, there is a 132\% improvement in throughput when the IBFD mode is activated. The reason behind the difference in improvement is that the amount of transmitted data during each transmission opportunity in the HD mode is either $\text{MPDU}_{\text{max}}$ (AP transmission) or $0.3 \times \text{MPDU}_{\text{max}}$ (STA transmission). Therefore, when the number of nodes is high, there is less likelihood that the AP transmits its larger load, which yields significantly lower throughput in the HD mode. On the other hand, each transmission opportunity in the IBFD mode results in transmitting an $\text{MPDU}_{\text{max}}$ in the DL direction \textit{and} $0.3 \times \text{MPDU}_{\text{max}}$ in the UL direction. When IBFD dual-frame aggregation is enabled, $\rho^{\textit{\tiny dual}}_{\textit{\tiny new}}$ becomes $0.6$. In this case, an increase of 23\% is consistently realized in throughput compared to the case of IBFD without aggregation. The throughput is increased by 46\% in each simulated case of $n$ when IBFD multi-frame aggregation is employed in the network, which corresponds to $\rho^{\textit{\tiny multi}}_{\textit{\tiny new}}=0.9$. The superior performance of IBFD multi-frame aggregation is expected since there is more data pushed in the UL direction. The upper limit for IBFD mode is indicated by the case when $\rho = 1$, and the increase in throughput from the case of IBFD without aggregation is 54\%.

\begin{figure}[!b]
\centering
\includegraphics[width=\columnwidth,trim=1.4in 3.3in 1.7in 3.5in,clip=true]
{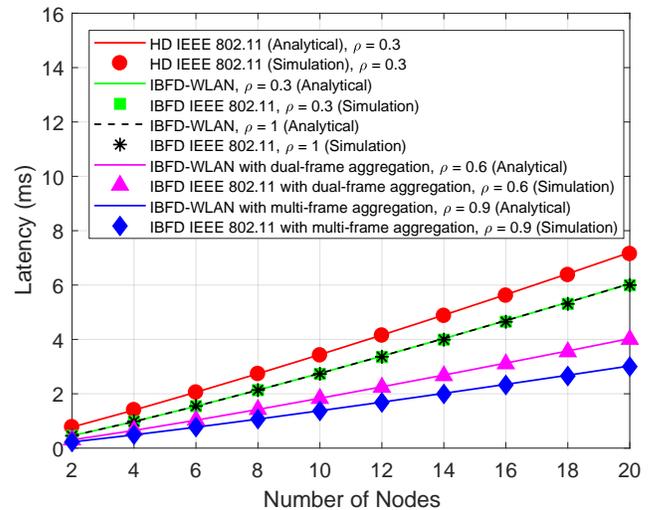}
\caption{Latency versus number of nodes $(\rho=0.3)$}
\label{murad6}
\end{figure}
Fig. \ref{murad6} displays the results for latency versus $n$. HD IEEE 802.11 exhibits the highest latency since in each transmission opportunity, either a DL or a UL frame is transmitted. Once IBFD transmission is implemented, there is reduction in latency since a DL frame \textit{and} a UL frame are delivered in each transmission. When $n=2$, there is a decrease of 42\% in latency compared to HD IEEE 802.11. When $n$ increases to 20, the reduction in latency is only 16\% since there are more frames at silent STAs experiencing delay while an IBFD transmission takes place between the AP and an STA. The case for IBFD IEEE 802.11 with $\rho = 1$ (maximum traffic symmetry) is plotted. However, no further improvement in latency is realized since the number of delivered frames in each transmission is still 2. When IBFD IEEE 802.11 is augmented by dual-frame aggregation, there is a 33\% improvement in latency for each case of $n$ compared to IBFD without aggregation. The reason is that in each transmission, 1 DL frame and 2 UL frames are delivered. When multi-frame aggregation is introduced, latency is reduced by 50\% compared to IBFD without aggregation since 1 DL frame and 3 UL frames are now served during each transmission. Clearly, aggregation is necessary to improve latency in IBFD-WLAN, and original $\rho$ values do not affect average latency, which decreases as a result of increasing the number of transmitted frames.

Constant $\rho$ values are assumed in this scenario, and the analytical values for $\eta$ are as calculated in TABLE \ref{murad.t2}. The analytical results simply match the simulated results as expected since deterministic $\rho$ values are assumed. It is worth noting that $\eta$ is not affected by the number of nodes in the network since only the size of useful traffic is relevant here. The results for $\eta$ indicate how well the channel is utilized in the assumed IBFD-WLAN network. IBFD link utilization becomes more sophisticated in the next section for random $\rho$ values. 
\begin{table} [!b]
\caption{IBFD link utilization for deterministic $\rho$ values.}
\label{murad.t2}
\centering
\begin{tabular}{|c||c|c|c|c|c|}
\hline
Aggregation Mode & $\rho_{\textit{\tiny original}}$ & $\gamma$ &$\rho_{\textit{\tiny new}}$ & $\Phi$ & $\eta$ \\\hline
Pure IBFD (no aggregation) & 0.3 & 1 & 0.3 & 0.3 & 65\% \\\hline
IBFD dual-frame aggregation & 0.3 & 2 & 0.6 & 0.6 & 80\% \\\hline
IBFD multi-frame aggregation & 0.3 & 3 & 0.9 & 0.9 & 95\% \\\hline
\end{tabular}
\end{table}

\begin{table}[!b]
\caption{IBFD frame aggregation rules for random $\rho$ values.}
\label{murad.t3}
\centering
\begin{tabular}{|c||c|c||c|c|}
\hline
\multirow{2}{*}{$\rho_{\textit{\tiny current}}$} & \multicolumn{2}{c||}{Dual-Frame} & \multicolumn{2}{c|}{Multi-Frame} \\\cline{2-5}
&$\gamma$ & $\rho_{\textit{\tiny new}}$ & $\gamma$ & $\rho_{\textit{\tiny new}}$ \\\hline
0.1 & 2 & 0.2 & 10 & 1 \\\hline
0.2 & 2 & 0.4 & 5 & 1 \\\hline
0.3 & 2 & 0.6 & 3 & 0.9 \\\hline
0.4 & 2 & 0.8 & 2 & 0.8 \\\hline
0.5 & 2 & 1.0 & 2 & 1 \\\hline
0.6 & 1 & 0.6 & 1 & 0.6 \\\hline
0.7 & 1 & 0.7 & 1 & 0.7 \\\hline
0.8 & 1 & 0.8 & 1 & 0.8 \\\hline
0.9 & 1 & 0.9 & 1 & 0.9 \\\hline
\end{tabular}
\end{table}

\subsection{Random $\rho$ Values}
\begin{figure}[!t]
\centering
\includegraphics[width=\columnwidth,trim=1.4in 3.3in 1.7in 3.5in,clip=true]
{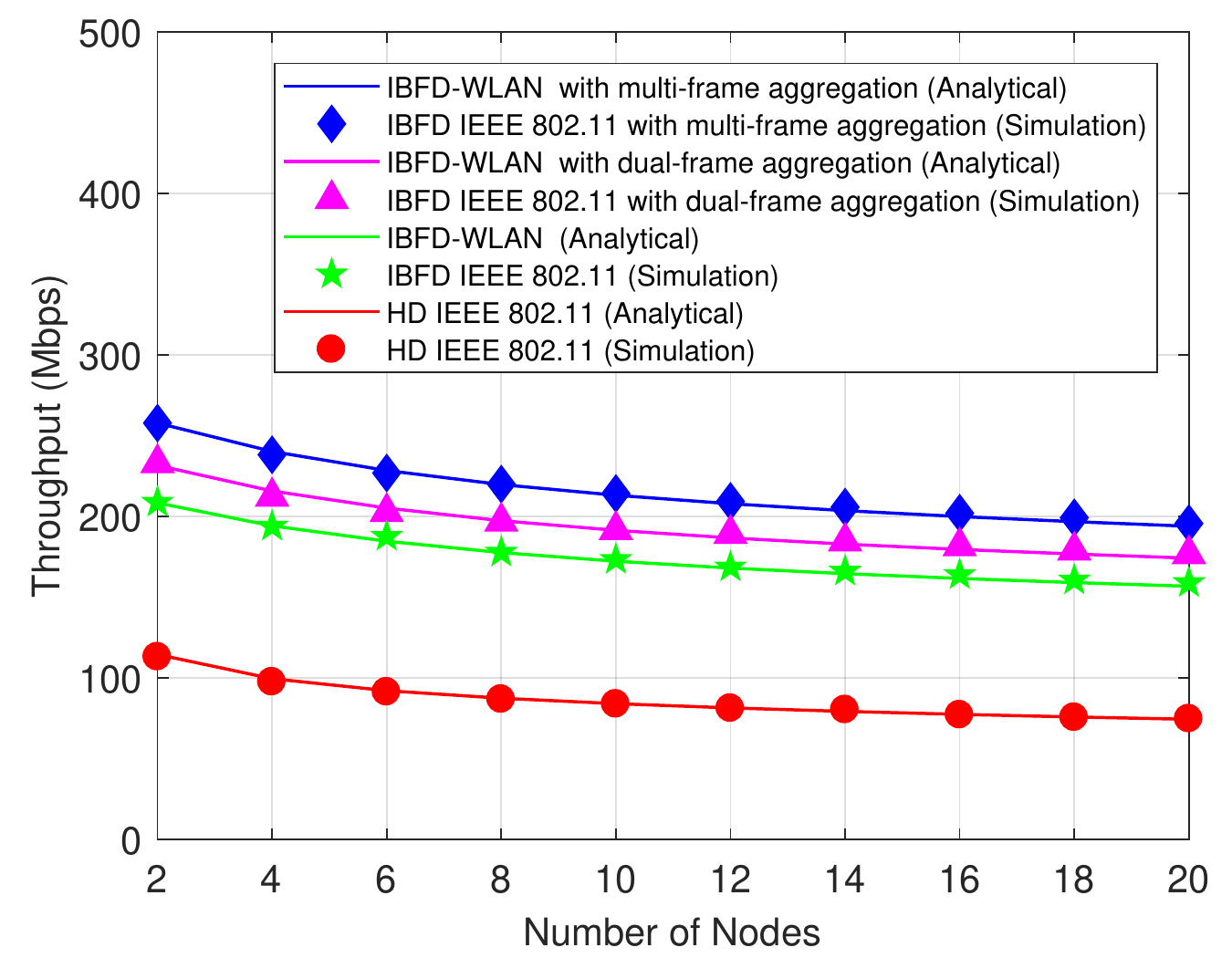}
\caption{Throughput versus number of nodes (random $\rho$ values, 200 runs)}
\label{murad7}
\end{figure}
In this section, a $\rho$ value for each client STA is randomly assigned such that $\rho$ is uniformly distributed over $\lbrace 0.1, 0.2, ..., 0.9 \rbrace$. New $\rho$ assignments are updated in each simulation run. The average result of 200 independent runs is reported for each simulation scenario. For IBFD dual-frame aggregation, only STAs with $0.1 \leq \rho \leq 0.5$ double their loads while the rest of STAs with $0.6 \leq \rho \leq 0.9$ maintain their original frames. When IBFD multi-frame aggregation is used, STAs with $0.6 \leq \rho \leq 0.9$ transmit their original loads while the rest of STAs aggregate their loads according to TABLE \ref{murad.t3}, which shows aggregation rules for both dual-frame and multi-frame modes. Fig. \ref{murad7} shows throughput results versus $n$. When $n=2$, IBFD introduces an 85\% increase in throughput compared to 112\% increase when $n=20$ (both comparisons are with the corresponding HD cases). The difference in improvement between the two cases is consistent with the case of deterministic $\rho$ values in section VI-B in that there is much less data transmitted in the DL direction when $n$ is high resulting in low throughput in the HD baseline case. When dual-frame aggregation is employed, there is an improvement of 11\% in throughput for each case of $n$ compared to the corresponding IBFD case without aggregation. For multi-frame aggregation, the increase in throughput is 24\%. Thus, IBFD multi-frame aggregation provides superior performance as expected since more UL transmission time is utilized.
\begin{figure}[!b]
\centering
\includegraphics[width=\columnwidth,trim=1.4in 3.3in 1.7in 3.5in,clip=true]
{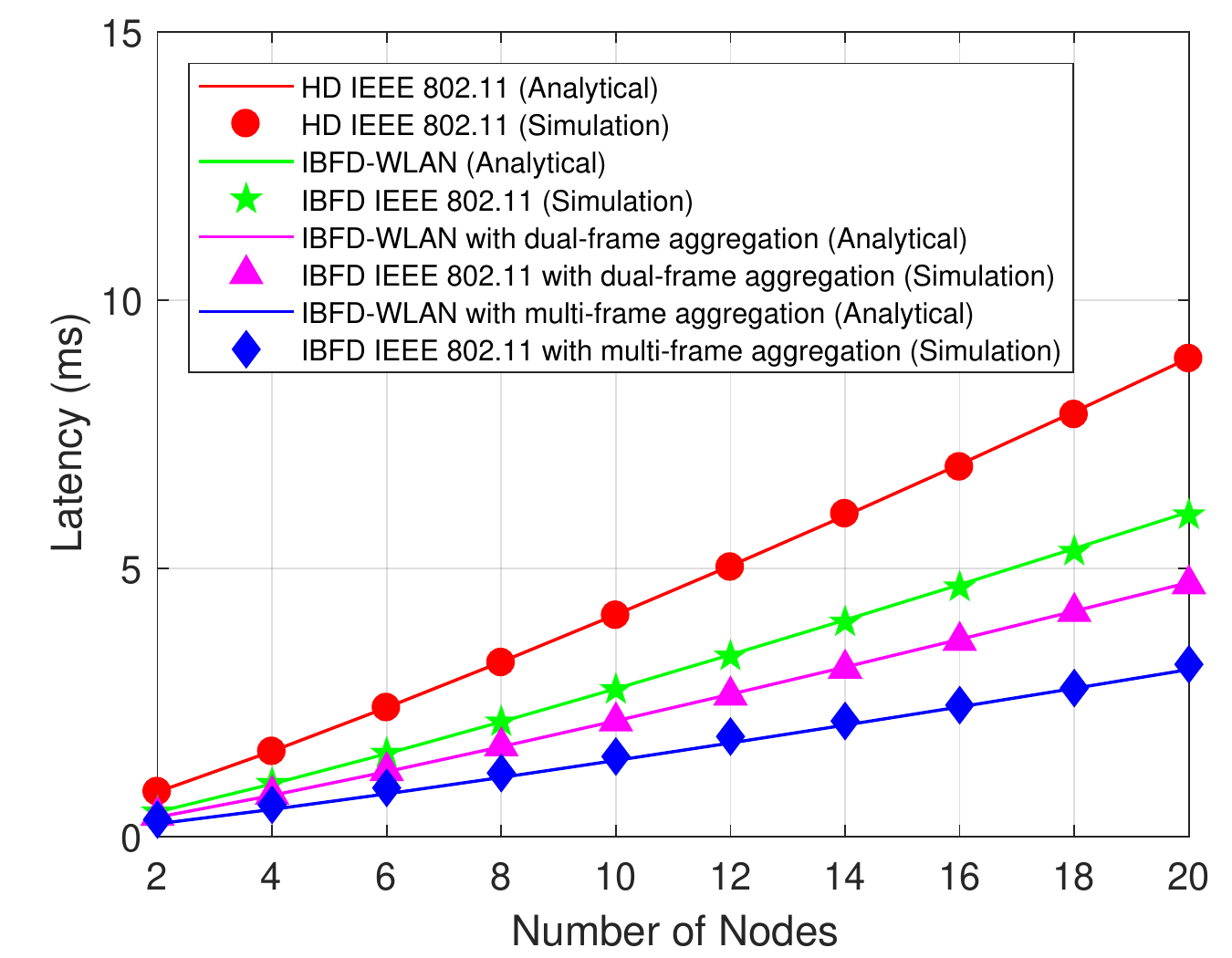}
\caption{Throughput versus number of nodes (random $\rho$ values, 200 runs)}
\label{murad8}
\end{figure}

Fig. \ref{murad8} displays latency results versus $n$ when $\rho$ values are random. IBFD without aggregation reduces latency by 45\% compared to the HD case when $n=2$, but the improvement in latency decreases as the number of nodes increases until it reaches 33\% when $n=12$. Even though increasing the number of nodes increases latency as expected, improvement in latency due to IBFD transmission remains unaffected and stays at 33\% as $n$ increases. This behavior is consistent with the scenario of deterministic $\rho$ values in that as $n$ continues to increase, nodes in the HD network experience higher delays while waiting for the active transmission to finish. Dual-frame aggregation introduces 16\% of reduction in latency when $n=2$ compared to IBFD without aggregation, and the improvement saturates to 22\% as $n$ increases. Multi-frame aggregation initially introduces 31\% improvement when $n=2$ compared to IBFD without aggregation, and the improvement saturates to 47\% for higher values of $n$. In both aggregation schemes, more reduction in latency is noted as $n$ increases. This behavior can be explained by the fact that as the number of nodes increases, there are more STAs that can initially have low $\rho$ values, which enable them to aggregate and transmit more frames. Therefore, the expected total number of transmitted frames in the UL direction increases as $n$ increases, which further reduces the average latency.
\begin{table}[t]
\caption{IBFD link utilization for random $\rho$ values.}
\label{murad.t4}
\centering
\begin{tabular}{|>{\centering\arraybackslash}p{4.9em}||>{\centering\arraybackslash}m{2.25em}|>{\centering\arraybackslash}m{2.25em}|>{\centering\arraybackslash}m{2.45em}||>{\centering\arraybackslash}m{2.25em}|>{\centering\arraybackslash}m{2.25em}|>{\centering\arraybackslash}m{2.45em}|}
\hline
\multirow{2}{*}{\makecell{Aggregation\\Mode}} & \multicolumn{3}{c||}{Analytical} & \multicolumn{3}{c|}{Simulation} \\\cline{2-7}
& $E[\gamma]$ & $\Phi$ & $\eta$& $E[\gamma]$ & $\Phi$ & $\eta$ \\\hline
None & 1 & 0.5000 & 75.00\% & 1 & 0.5044 & 75.22\% \\\hline
Dual-frame & 1.5556 & 0.6667 & 83.34\% & 1.5608 & 0.6649 & 83.24\% \\\hline
Multi-frame & 2.8889 & 0.8556 & 92.78\% & 2.8915 & 0.8545 & 92.72\% \\\hline
\end{tabular}
\end{table}

Since the value of $\rho$ is equally likely to be one of the uniformally distributed values between 0.1 and 0.9, $E[\gamma]$ can be directly calculated based on the values in TABLE \ref{murad.t3}. In addition, analytical values for $\eta$ in the case of random $\rho$ values are readily obtained based on the calculated values of $\Phi$, and the results are summarized in TABLE \ref{murad.t4}. Average simulation results from 200 runs are also reported for $E[\gamma]$, $\Phi$, and $\eta$. It is noted that the values of $\eta$ are directly proportional to the values of $\Phi$ as expected. When random $\rho$ values are introduced into the system, simulation results for IBFD performance metrics are in strong agreement with the analytical results and still consistent with the case of deterministic $\rho$ values.

\section{Conclusion}
In this paper, an accurate model characterizing IEEE 802.11 DCF for IBFD-WLAN is presented. The model is based on a two-dimensional DTMC framework. The concepts of IBFD transmission and frame aggregation are combined to maximize throughput and minimize latency in WLANs. The proposed aggregation schemes increase the utilization of available UL transmission time that would otherwise be unused. Each client STA uses its own traffic information to make a localized decision about the option and size of aggregation. Aggregation is necessary to minimize latency in IBFD-WLANs. The proposed analytical model and related metrics are robust and produce values coinciding with the simulated results even when randomness is introduced in the system. Since no changes to IEEE 802.11 protocol were introduced in this paper, the proposed IBFD frame aggregation schemes would be backward compatible with future IEEE 802.11 releases. Network throughput, average latency, and link utilization are proposed as metrics to quantify potential enhancements resulting from introducing IBFD in WLANs. 

\appendices
\setcounter{equation}{0}
\renewcommand{\theequation}{A.\arabic{equation}}
\section{Derivations of $b_{i,0}$ and $b_{0,0}$}
Start with calculating $b_{1,0}$ in terms of $b_{0,0}$ based on Fig. \ref{murad2}
\begin{equation}
\begin{split}
b_{1,0} &= b_{0,0}\frac{p}{W_1}+b_{1,1}\cdot\alpha\\
&= b_{0,0}  \frac{p}{W_1}   +  \alpha \cdot (b_{0,0} \frac{p}{W_1} + \alpha \cdot b_{1,2})\\
&=  b_{0,0} \frac{p}{W_1} + \alpha \cdot b_{0,0} \frac{p}{W_1} + \alpha^2 \cdot b_{1,2}\\
&=  b_{0,0}  \frac{p}{W_1}(1+\alpha+\alpha^2+...+\alpha^{W_1-2})+\alpha^{W_1-1} \cdot  b_{1,W_1-1}\\
&=  b_{0,0}  \frac{p}{W_1}(1+\alpha+\alpha^2+...+\alpha^{W_1-2})+\alpha^{W_1-1} \cdot  b_{0,0} \frac{p}{W_1}\\
&=  \frac{b_{0,0}  \cdot p}{W_1} \sum_{j_1=0}^{W_1-1} \alpha^{j_1}\\
&=  \frac{b_{0,0} \cdot p}{W_1} \cdot \frac{1-\alpha^{W_1}}{1-\alpha} 
\end{split}
\label{b1_0_app}
\end{equation}
based on resolving the sum of the geometric series. Similarly, calculate $b_{2,0}$ and substitute for $b_{1,0}$ from (\ref{b1_0_app})
\begin{equation}
\begin{split}
b_{2,0} &= \frac{b_{1,0} \cdot p}{W_2}\sum_{j_2=0}^{W_2-1} \alpha^{j_2} = \frac{b_{0,0} \cdot p}{W_1} \cdot \frac{1-\alpha^{W_1}}{1-\alpha} \cdot \frac{p}{W_2} \cdot \frac{1-\alpha^{W_2}}{1-\alpha}\\
&=\frac{b_{0,0} \cdot p^2}{W_1 \cdot W_2} \cdot \frac{(1-\alpha^{W_1})(1-\alpha^{W_2})}{(1-\alpha)^2}.
\end{split}
\end{equation}
Noticing the pattern in $b_{1,0}$ and $b_{2,0}$, $b_{i,0}$ can be written as
\begin{equation}
\Rightarrow b_{i,0} = b_{0,0} (\frac{p}{1-\alpha})^i \prod_{j=1}^{i} \frac{1-\alpha^{W_j}}{W_j}, 1 \leq i \leq m.
\label{bi_0_app}
\end{equation}
For $b_{0,0}$, it can directly be deduced from Fig. \ref{murad2} that 
\begin{equation}
\begin{split}
b_{0,0} &= \sum_{i=0}^{m} b_{i,0} \cdot (1-p) \cdot \frac{1}{W_0} + \sum_{i=0}^{m} \sum_{k=1}^{W_i - 1} b_{i,k} \cdot \beta \cdot \frac{1}{W_0}\\
&+ b_{m,0} \cdot p \cdot \frac{1}{W_0} + \alpha \cdot b_{0,1}\\
&= \tau \cdot (1-p) \cdot \frac{1}{W_0} + (1-\tau) \cdot \beta \cdot \frac{1}{W_0}\\
&+ b_{m,0} \cdot p \cdot \frac{1}{W_0} + \alpha \cdot b_{0,1}\\
&= \underbrace{\frac{\tau (\alpha - p) + 1-\alpha + p \cdot b_{m,0}}{W_0}}_{Z} + \alpha \cdot b_{0,1}\\
&= Z + \alpha \cdot (Z + \alpha \cdot b_{0,2}) = Z (1+\alpha)+ \alpha^2 \cdot b_{0,2}\\
&= Z(1+\alpha+\alpha^2+...+\alpha^{W_0-2})+\alpha^{W_0-1} \cdot  \underbrace{b_{0,W_0-1}}_{Z}\\
&= Z \sum_{j=0}^{W_0-1} \alpha^j = Z \cdot \frac{1-\alpha^{W_0}}{1-\alpha}\\
&= \frac{1-\alpha^{W_0}}{1-\alpha} \frac{\alpha - p}{W_0} \cdot \tau + \frac{1-\alpha^{W_0}}{W_0} + \frac{1-\alpha^{W_0}}{1-\alpha} \frac{p \cdot b_{m,0}}{W_0}.
\end{split}
\label{b0_0_app}
\end{equation}
By substituting the expression for $b_{m,0}$ from (\ref{bi_0_app}) in (\ref{b0_0_app}), $b_{0,0}$ becomes
\begin{equation}
\begin{split}
b_{0,0} &= \frac{1-\alpha^{W_0}}{W_0}\Bigg[ \frac{\alpha - p}{1- \alpha} \cdot \tau + 1\Bigg]\\
&+ \frac{1-\alpha^{W_0}}{(1- \alpha) W_0} \cdot p \cdot b_{0,0} (\frac{p}{1-\alpha})^m \prod_{j=1}^{m} \frac{1-\alpha^{W_j}}{W_j}
\end{split}
\end{equation}
\begin{equation}
\Rightarrow b_{0,0} = \frac{\frac{1-\alpha^{W_0}}{W_0}\Bigg[\frac{\alpha - p}{1- \alpha} \cdot \tau + 1\Bigg]}{1-(\frac{p}{1-\alpha})^{m+1} \displaystyle \prod_{j=0}^{m} \frac{1-\alpha^{W_j}}{W_j}}.
\end{equation}

\section{Derivation of $P^\textit{\tiny IBFD}_{s}$}
\setcounter{equation}{0}
\renewcommand{\theequation}{B.\arabic{equation}}
Based on the four cases of the conditional probabilities given in section IV-D for $P^\textit{\tiny IBFD}_{s}$
\begin{equation}
\begin{split}
&P^\textit{\tiny IBFD}_{s} = \frac{\tau_{_{AP}}(1-\tau_{_{STA}})^{n-1}}{P^\textit{\tiny IBFD}_{tr}}\\
&+\frac{(n-1)\tau_{_{STA}}(1-\tau_{_{AP}})(1-\tau_{_{STA}})^{n-2}}{P^\textit{\tiny IBFD}_{tr}}\\
&+\frac{1}{n} \cdot \frac{1}{n-1} \cdot \frac{\tau_{_{AP}}\tau_{_{STA}}(1-\tau_{_{STA}})^{n-2}}{P^\textit{\tiny IBFD}_{tr}}\\
&+\frac{n-1}{n} \cdot \frac{1}{n-1} \cdot \frac{\tau_{_{AP}}\tau_{_{STA}}(1-\tau_{_{STA}})^{n-2}}{P^\textit{\tiny IBFD}_{tr}}\\
&=\frac{\tau_{_{AP}}(\overline{\tau}_{_{STA}})^{n-1}}{P^\textit{\tiny IBFD}_{tr}}\\
&+\frac{(n-1)\tau_{_{STA}}(\overline{\tau}_{_{AP}})(\overline{\tau}_{_{STA}})^{n-2}}{P^\textit{\tiny IBFD}_{tr}}\\
&+\frac{1}{n(n-1)} \cdot \frac{\tau_{_{AP}}\tau_{_{STA}}(\overline{\tau}_{_{STA}})^{n-2}}{P^\textit{\tiny IBFD}_{tr}}\\
&+\frac{1}{n} \cdot \frac{\tau_{_{AP}}\tau_{_{STA}}(\overline{\tau}_{_{STA}})^{n-2}}{P^\textit{\tiny IBFD}_{tr}}.
\end{split}
\end{equation}
According to (\ref{P_tr_IBFD}) in section V-A,
\begin{equation}
\begin{split}
&P^\textit{\tiny IBFD}_{tr} = 1-[(1-\tau_{_{AP}}) (1-\tau_{_{STA}})^{n-1}]\\
&= 1-[(\overline{\tau}_{_{AP}}) (\overline{\tau}_{_{STA}})^{n-1}]
\end{split}
\end{equation}
\begin{equation}
\begin{split}
\\
&\Rightarrow P^\textit{\tiny IBFD}_{s} =\\
\\
&\frac{\tau_{_{AP}}(\overline{\tau}_{_{STA}})^{n-1}+(n-1)\tau_{_{STA}}(\overline{\tau}_{_{AP}})(\overline{\tau}_{_{STA}})^{n-2}}{1-[(\overline{\tau}_{_{AP}}) (\overline{\tau}_{_{STA}})^{n-1}]}\\
\\
&+\frac{\tau_{_{AP}}\tau_{_{STA}}(\overline{\tau}_{_{STA}})^{n-2}}{(n-1)\big\{1-[(\overline{\tau}_{_{AP}}) (\overline{\tau}_{_{STA}})^{n-1}]\big\}}.
\end{split}
\end{equation}

\ifCLASSOPTIONcaptionsoff
  \newpage
\fi

\bibliographystyle{IEEEtran.bst}

\vskip 0pt plus -1fil



\begin{IEEEbiography}
[{\includegraphics[width=1in,height=1.25in,clip,keepaspectratio]{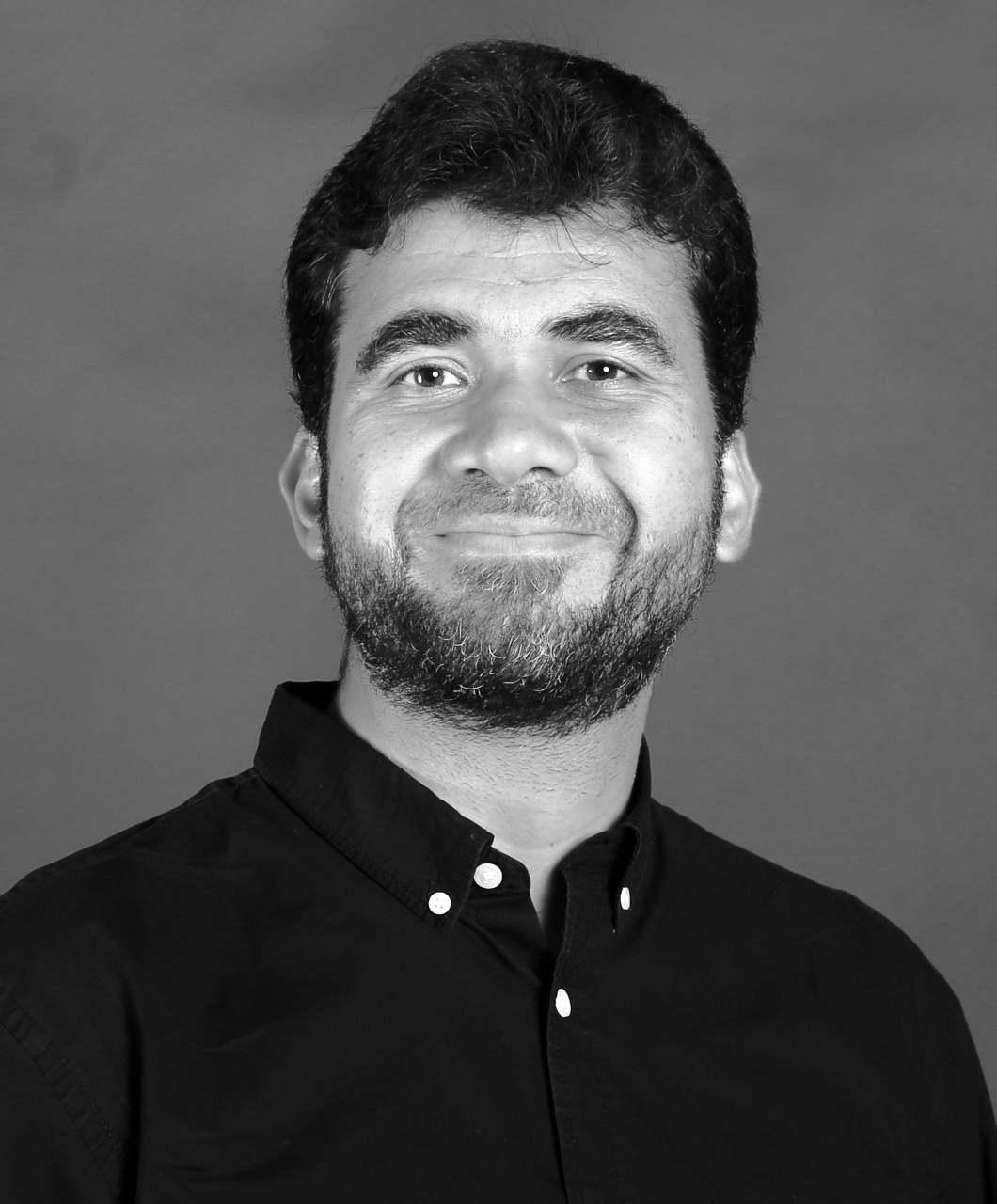}}]
{Murad Murad} received the Bachelor of Science degree (Summa Cum Laude) from California State University, Long Beach and the Master's of Engineering degree from The University of British Columbia in Vancouver, Canada. He is a PhD candidate in electrical engineering at University of California, Irvine. As a research associate in Wireless Systems and Circuits Laboratory, Murad pursues original contributions to achieve cutting-edge results in improving the design of wireless networks based on In-Band Full-Duplex techniques. Furthermore, he completed certification programs in higher education pedagogy. Murad previously had supervisory experience in leading Information Technology teams to run operations related to telecommunications and data networking. He also had technical experience as a communications engineer working on communications systems and projects. Murad received training on various communications and networking systems from leading vendors in the industry.
\end{IEEEbiography}
\begin{IEEEbiography}
[{\includegraphics[width=1in,height=1.25in,clip,keepaspectratio]{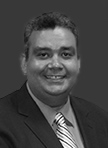}}]
{Ahmed M. Eltawil} (S'97-M'03-SM'14) is a Professor at the University of California, Irvine. He has been with the Department of Electrical Engineering and Computer Science since 2005 where he is the founder and director of the Wireless Systems and Circuits Laboratory. His current research interests are in the general area of low power digital circuit and signal processing architectures with an emphasis on mobile computing and communication systems. In addition to his department affiliation, he is also affiliated to a number of research centers across the University of California, Irvine. He received the Doctorate degree from the University of California, Los Angeles, in 2003 and the M.Sc. and B.Sc. degrees (with honors) from Cairo University, Giza, Egypt, in 1999 and 1997, respectively. Dr. Eltawil has been on the technical program committees and steering committees for numerous workshops, symposia, and conferences in the areas of low power computing and wireless communication system design. He received several awards, as well as distinguished grants, including the NSF CAREER grant in 2010 supporting his research in low power systems. In 2015, Dr Eltawil founded Lextrum Inc., to develop and commercialize full duplex solutions for 5G communications systems.
\end{IEEEbiography}
\end{document}